%Paper: hep-th/9411177
%From: Christian Schubert <schubert@hades.ifh.de>
%Date: Wed, 23 Nov 1994 22:22:18 +0100 (MET)

%%%%%%%%%%%%%%%%%%%%%%%%%%%%%%%%%%%%%%%%%%%%%%%%%%%%%%
%%%%%%%%%%%%%%%%%%%%%%%%%%%%%%%%%%%%%%%%%%%%%%%%%%%%%%
%%%%%               PLAIN TEX                   %%%%%%
%%%%%%%%%%%%%%%%%%%%%%%%%%%%%%%%%%%%%%%%%%%%%%%%%%%%%%
%%%%%%%%%%%%%%%%%%%%%%%%%%%%%%%%%%%%%%%%%%%%%%%%%%%%%%
\magnification=\magstep1
\baselineskip=12.67pt plus 2pt minus 1 pt %10pt, 18pt
\hsize=6truein
\hoffset=10pt
\voffset=20pt
\parskip=1pt plus 1pt
\vsize=8truein
\footline{\ifnum\pageno=1{\hss}\else\hss\tenrm\folio\hss\fi}
\hfuzz=1pt
\tolerance=10000
\font\abfnt=cmbx10
\font\reffnt=cmbx9 scaled \magstep2
%\font\title=cmbx12
\font\title=cmr10 scaled \magstep3
\font\sc=cmcsc10
\font\brm=cmr9 scaled 833
\font\bsl=cmsl9 scaled 833
\newif\ifninebanner\ninebannertrue
\newdimen\bannerskip
\def\beginflushfootnote{\bgroup\parindent=0pt\footnote}
 \def\endflushfootnote{\egroup}
%
%
%%%%%%%%%%%%%%%%%% own definitions
\def\n{\noindent}
\def\del{\partial}
\def\Vka{V_{\kappa}}
\def\Vla{V_{\lambda}}
\def\Vmu{V_{\mu}}
\def\Vnu{V_{\nu}}
\def\Vro{V_{\rho}}
\def\Vkala{V_{\kappa\lambda}}
\def\Vkamu{V_{\kappa\mu}}
\def\Vkanu{V_{\kappa\nu}}
\def\Vlamu{V_{\lambda\mu}}
\def\Vlanu{V_{\lambda\nu}}
\def\Vmunu{V_{\mu\nu}}
\def\Vnuro{V_{\nu\rho}}
\def\Vkalamu{V_{\kappa\lambda\mu}}
\def\Vkalanu{V_{\kappa\lambda\nu}}
\def\Vkalaro{V_{\kappa\lambda\rho}}
\def\Vkamunu{V_{\kappa\mu\nu}}
\def\Vlamunu{V_{\lambda\mu\nu}}
\def\Vmunuro{V_{\mu\nu\rho}}
\def\Vkalamunu{V_{\kappa\lambda\mu\nu}}
\def\Vkalamuro{V_{\kappa\lambda\mu\rho}}
\def\Vkalamunuro{V_{\kappa\lambda\mu\nu\rho}}
\def\Vkalamunurosi{V_{\kappa\lambda\mu\nu\rho\sigma}}
%%%%%%%%%%%%%%%%%%
\line{\hfill DESY--94--221\phantom{xxx}}
\line{\hfill HD--THEP--94--26} \vskip60pt
\centerline{\title An Improved Heat Kernel Expansion }
\centerline{\title From Wordline Path Integrals}
\vskip12pt
\centerline{\sc D.~Fliegner, P.~Haberl, M.G.~Schmidt}
\smallskip
\centerline{\sl Institut f\"ur Theoretische Physik}
\centerline{\sl Universit\"at Heidelberg}
\centerline{\sl Philosophenweg 16}
\centerline{\sl D-69120 Heidelberg (Germany)}
\smallskip\smallskip
\centerline{\sc C.~Schubert}\smallskip
\centerline{\sl Institut f\"ur Hochenergiephysik Zeuthen}
\centerline{\sl Deutsches Elektronen-Synchrotron DESY}
\centerline{\sl Platanenallee 6}
\centerline{\sl D-15738 Zeuthen (Germany)}
\vskip40pt
\beginflushfootnote{}{{\baselineskip\bannerskip \brm\noindent
Talk given by P.~Haberl at the Conference
{\bsl Heat Kernel Techniques and Quantum Gravity},
University of Manitoba, Winnipeg (Canada), Aug 2-6 1994
\par}}\endflushfootnote
\centerline{\abfnt Abstract}
\vskip10pt
\noindent
The one--loop effective action for the case of a massive
scalar loop in the background of both a scalar potential
and an abelian or non--abelian gauge field is written in
a one--dimensional path integral representation. From this
the inverse mass expansion is obtained by Wick contractions
using a suitable Green function, which allows the computation
of higher order coefficients.
For the scalar case, explicit results are presented
up to order ${\cal O}(T^8)$
in the proper time expansion.
The relation to previous work is clarified.
\vfill\eject
The calculation of one--loop effective actions as
determinants of certain operators $M$ has a long history,
and numerous representations in closed form or as
expansions are available [1]. A common starting point is
to write the one--loop determinant in the Schwinger proper
time representation,
$$
\Gamma_{\rm eff}=-{\rm ln}({\rm det M})=-{\rm Tr}({\rm ln} M)
=\int_0^\infty{{\rm d}T\over T}{\rm Tr}\;{\rm e}^{-T M}\;,
\eqno(1)$$
where a suitable regularization procedure is implicitly
understood. In the conventional approach, the operator
trace is written as the diagonal element of the heat kernel,
which is then evaluated using recursive or nonrecursive
methods.
\par
The terms of an expansion of the effective action can be
grouped in various ways, leading to different approximations.
For instance, Schwinger--type formulas [2] are obtained from
terms where a fixed number of derivatives act on an arbitrary
number of external fields. Conversely, all terms with a fixed
number of external fields but arbitrarily many derivatives
lead to the Barvinsky--Vilkovisky form factors [3].
In applications to physics one is often interested in an
effective theory obtained from integrating out heavy
degrees of freedom. In this case, the relevant contributions
to low energy physics are the first terms of an inverse mass
expansion. In the following we will consider this approach,
which results from expanding the exponential in eq.\ (1) in
powers of the proper time $T$.
\par
In recent years there has been growing interest in the
explicit form of higher order coefficients of the inverse mass
expansion, especially from calculations of quark determinants
in QCD [4] and from considerations on the electroweak phase
transition. In the calculation of the fluctuation determinant
around a sphaleron configuration [5], it was observed that the
slow convergence of the series made it difficult to arrive at
quantitative statements, and that therefore it would be
desirable to push the expansion as high as possible.
This has also been the original motivation for our work.
\par
Our main emphasis is therefore on the actual computation of
higher order coefficients in the inverse mass expansion. With
conventional techniques this turns out to be a tedious task,
with the highest results known to us being of order
${\cal O}(T^6)$ for the ungauged theory [6] and
${\cal O}(T^5)$ for the gauged case [7].
\par
The method we will discuss in the following makes
use of recent progress in calculating one--loop amplitudes
in field theory. In the basic work of Bern and Kosower [8],
ordinary amplitudes were represented as superstring amplitudes
in the limit of infinite string tension. This leads to new rules
for evaluating one--loop diagrams which are in many aspects
superior to conventional methods. For instance, making use of
certain supersymmetry transformations allows to derive simple
relations between scalar and spinor loops. Later, Strassler [9]
showed that -- at least for the cases we will consider -- the
set of one--loop Bern--Kosower rules can equivalently be
obtained from one--dimensional path integrals.
\par
It is well known that one--loop effective actions can at least
formally be written as one--dimensional path integrals [10].
The new aspect of the present work is the actual evaluation of
the path integral, applying the techniques of Bern, Kosower and
Strassler. The calculations turn out to be more efficient than
conventional heat kernel methods. Among the advantages are the
small number of terms produced in intermediate steps and that the
results are obtained automatically in a ``minimal'' basis of
effective operators, without further use of partial integrations.
Finally, the method is well suited to computerization and allowed
us to push the inverse mass expansion to ${\cal O}(T^{11})$ for
the ungauged theory.
\par
%\vfill\eject
Let us first discuss the case of a massive scalar loop in the
background of both a gauge field (abelian or non--abelian) and a
generally matrix--valued scalar potential, for which the
fluctuation operator $M$ takes the form
$$
M = - D^2 + m^2 + V(x) \qquad{\rm with}\qquad
D_\mu=\del_\mu - ig A_\mu \,.
\eqno(2)$$
The following calculations are performed in $d$--dimensional
euclidean space-time. Operator Traces will be denoted by
${\rm Tr}$, while ordinary matrix traces are written as
${\rm tr}$. Starting with the proper time integral
$$
\Gamma [A,V]
= \int_0^\infty {{\rm d}T \over T} {\rm e}^{-m^2 T} {\rm Tr}
\exp [-T(-D^2 + V)]
\eqno(3)$$
one evaluates the trace in $x$-space and decomposes the interval
$T$ in the exponent, yielding ($x_N=x_0=x$):
$$ \eqalign{
\Gamma [A,V]
&= \int_0^\infty {{\rm d}T \over T} {\rm e}^{-m^2 T} {\rm tr}
\int \langle x | \exp [-T(-D^2+V)] | x \rangle \cr
&= \int_0^\infty {{\rm d}T \over T} {\rm e}^{-m^2 T} {\rm tr}
\int {\prod_{i=0}^N} {\rm d}x_i \langle x_i | \exp \Bigl[
- {T\over N} (-D^2+V)\Bigr]| x_{i-1} \rangle \cr  }
\eqno(4)$$
Inserting complete sets of momentum states
$\int|p_i\rangle\langle p_i|$ and performing the usual Gaussian
integration we obtain in the limit of $N\to\infty$ a
representation of $\Gamma[A,V]$ as a one--dimensional path
integral on the space of closed loops in space-time with fixed
circumference $T$,
$$
\Gamma[A,V]=
\int_0^\infty {{\rm d}T\over T} {\rm e}^{-m^2T} {\rm tr} \;\;
{\cal P}\!\! \int_{x(T)=x(0)} \!\!\!\!\!\!\!\!\!\!\!\! {\cal D}x
\exp \Bigl[ - \int_0^T {\rm d}\tau \bigl( {\dot{x}^2\over 4}
+ ig \dot{x}^\mu A_\mu  +V(x) \bigr)\Bigr]\,,
\eqno(5)$$
where ${\cal P}$ denotes path ordering.
The worldline path integral is then evaluated by Wick
contractions using the Green function of the Laplacian on the
circle with periodic boundary conditions [10,9,11].
Since the defining equation
$$
{1\over 2} {\del^2\over\del\tau_1^2} G_B(\tau_1,\tau_2)
= \delta(\tau_1-\tau_2)
\eqno(6)$$
has no solution -- similar to the case of the Poisson equation
for a charge in compact space -- one has to add a
``background charge'' on the worldline. With a uniform background
charge $\rho$ eq.\ (6) takes the form
$$
{1\over 2} {\del^2\over\del\tau_1^2} G_B(\tau_1,\tau_2)
= \delta(\tau_1-\tau_2) - \rho(\tau_1)
= \delta(\tau_1-\tau_2) - {1\over T} \;,
\eqno(7)$$
with the solution
$$\eqalign{
G_B(\tau_1,\tau_2) &=|\tau_1-\tau_2|-{(\tau_1-\tau_2)^2\over T}
\;, \cr \dot{G}_B(\tau_1,\tau_2) &= {\rm sign}(\tau_1-\tau_2)
- {2\over T} (\tau_1-\tau_2)  \;. \cr}
\eqno(8)$$
Elementary fields are thus contracted using the following rules:
$$\eqalign{
\langle x^{\mu}(\tau_1) x^{\nu}(\tau_2) \rangle &=
-g^{\mu\nu} G_B(\tau_1,\tau_2) =
-g^{\mu\nu} \Bigl[|\tau_1-\tau_2|-{(\tau_1-\tau_2)^2\over T}
\Bigr] \;,\cr \langle \dot{x}^{\mu}(\tau_1) x^{\nu}(\tau_2)
\rangle &=-g^{\mu\nu} \dot{G}_B(\tau_1,\tau_2) = -g^{\mu\nu}
\Bigl[ {\rm sign}(\tau_1-\tau_2)-{2(\tau_1-\tau_2)\over T}
\Bigr] \;,\cr
\langle \dot{x}^{\mu}(\tau_1) \dot{x}^{\nu}(\tau_2) \rangle &=
+g^{\mu\nu} \ddot{G}_B(\tau_1,\tau_2) =
+g^{\mu\nu} \Bigl[2\delta(\tau_1-\tau_2)-{2\over T}\Bigr]
\;. \cr }
\eqno(9)$$
\par
However, the introduction of a background charge causes a
problem: the Green function $G_B$ cannot be applied to the full
path integral directly, for partial integration on the circle
yields the identity
$$
\int_0^T {\rm d}\tau_2 \;{1\over 2} G_B(\tau_1,\tau_2)
\ddot{x}(\tau_2) = x(\tau_1) - {1\over T} \int_0^T
{\rm d}\tau_2 \;x(\tau_2)\,,
\eqno(10)$$
where the second term should vanish. One should therefore
integrate over relative loop coordinates
only, introducing a loop center of mass $x_0$,
$$
x^{\mu}(\tau) = x^\mu_0 + y^\mu(\tau)
\eqno(11)$$
with
$$
\int_0^T {\rm d}\tau \;y^{\mu}(\tau) = 0\,,
\eqno(12)$$
and extracting the integral over the center of mass from
the path integral:
$$
\int {\cal D}x = \int {\rm d}^dx_0 \int {\cal D}y \;.
\eqno(13)$$
The result is a representation of the effective Lagrangian as
an integral over the space of all loops with a common center of
mass $x_0$. From this we get the inverse mass expansion by
expanding the (path ordered) interaction exponential,
using $\dot{y}^\mu=\dot{x}^\mu$. If we take the background gauge
field to be in Fock-Schwinger gauge with respect to $x_0$,
$$
y^\mu A_\mu(x_0+y) \equiv 0
\quad {\rm for}\;\;{\rm all}\;\; y\,,
\eqno(14)$$
the gauge field may be written as
$$
A_\mu(x_0+y) = y^\rho \int_0^1 {\rm d}\eta \;\eta
F_{\rho\mu}(x_0+\eta y)
\eqno(15)$$
and $F_{\rho\mu}$ as well as $V$ can be Taylor-expanded
covariantly,
$$\eqalign{
F_{\rho\mu}(x_0+\eta y) &= {\rm e}^{\eta y D}
F_{\rho\mu}(x_0) \;,\cr
V(x_0+ y) &= {\rm e}^{ y D} V(x_0) \;.\cr  }
\eqno(16)$$
Thus there is a covariant Taylor expansion of the gauge
field $A$,
$$
A_\mu(x_0+y) = \int_0^1 {\rm d}\eta\;\eta \,y^\rho
{\rm e}^{\eta y D} F_{\rho\mu}(x_0) = {1\over 2} y^\rho
F_{\rho\mu} + {1\over 3} y^\nu y^\rho D_\nu F_{\rho\mu}
+ ...
\eqno(17)$$
Using the formulae above we obtain a manifestly covariant
expansion of eq.\ (5), namely
$$ \eqalign{
\Gamma[F,V] & =  \int_0^\infty {{\rm d}T\over T}
{\rm e}^{-m^2 T} \;{\rm tr}\int {\rm d}^d x_0 \sum_{n=0}^\infty
{(-1)^n\over n} T \int {\cal D}y \;{\rm exp} \Bigl[ -
\int_0^T \!\!\!{\rm d}\tau \;{\dot{y}^2\over 4} \Bigr] \cr
\times &\int_0^{\tau_1=T}\!\!\!{\rm d}\tau_2 \int_0^{\tau_2}
\!\!{\rm d}\tau_3 ... \int_0^{\tau_{n-1}} \!\!\!{\rm d}\tau_n
\prod_{j=1}^n \Bigl[\;{\rm e}^{ y(\tau_j)D_{(j)}} V^{(j)}(x_0)
\;+\cr
&\qquad\qquad+ \; ig \dot{y}^{\mu_j}(\tau_j)
y^{\rho_j}(\tau_j)\int_0^1 {\rm d}\eta_j \eta_j \;
 {\rm e}^{\eta_j y(\tau_j) D_{(j)}}
F_{\rho_j\mu_j}^{(j)}(x_0)  \Bigr] \;,\cr
}
\eqno(18)$$
where the first $\tau$-integration has been eliminated by
using the freedom of choosing the point $\tau =0$ somewhere
on the loop, which produces a factor of $T/n$.
The normalization is such that in the noninteracting case
$$
{\rm tr}\;{\cal P}\!\! \int {\cal D}y \exp \Bigl[ -\int_0^T
{\rm d} \tau \Bigl( {\dot{y}^2\over 4} \Bigr)\Bigr] =
(4\pi T)^{-d/2}\;.
\eqno(19)$$
\par
The calculation of the inverse mass expansion to some
fixed order $N$ requires the following steps:
\par
{\it (i) Wick-contractions\/}: The sum in eq.\ (18) has to be
truncated at $n=N$ and all possible Wick-contractions have
to be performed using the contraction rules for exponentials
well known from string theory,
$$\eqalign{
\langle {\rm e}^{y(\tau_1)\del_{(1)}}
{\rm e}^{y(\tau_2)\del_{(2)}} \rangle &=
{\rm e}^{- G(\tau_1,\tau_2) \del_{(1)} \del_{(2)}} \;,\cr
\langle \dot{y}^\mu(\tau_1) {\rm e}^{y(\tau_1)\del_{(1)}}
{\rm e}^{y(\tau_2)\del_{(2)}} \rangle &=
- \dot{G}_B(\tau_1,\tau_2) \del^\mu_{(2)}
{\rm e}^{-G_B(\tau_1,\tau_2)\del_{(1)}\del_{(2)}} \;,\cr }
\eqno(20)$$
and so on. In the covariant formulation the Wick-contractions
lead to ordered exponentials of the form
$$
{\rm exp} \Bigl[ - \sum_{i<k} \eta_i \eta_k G_B(\tau_i,
\tau_k) D^{(i)} D^{(k)} \Bigr]\,,
\eqno(21)$$
where the ordering of $D^{(i)}$'s in the exponential
series with different $i$'s corresponds to the ordering
of the $F^{(i)}$ and $V^{(i)}$, while any polynomial in
$D^{(i)}$ for fixed $i$ has to be written in all possible
orderings divided by the total number of orderings.
Additional factors of $D^{(i)}$ have to be handled in
the same way. Finally, the ordered exponentials have
to be expanded to the required order. Alternatively
one might use the explicit Taylor-expansions of $F$ and $V$
where one would be left with a much larger number of
Wick-contractions of polynomials using the formulae (9).
\par
{\it (ii) Integrations\/}: The next step is to perform the
$\tau$- and $\eta$-integrations, which are polynomial.
The integrands of the $\tau$-integrations consist of
the worldline Green function $G_B$ and its first and
second derivative. Using the scaling properties of these
functions, a rescaling to the unit circle is possible.
% The covariant formulation turns out to have the further
% advantage that all integrals involving at least one
% $\delta$-function from $\ddot{G}_B$ do vanish. Therefore
% $\ddot{G}_B$ can simply replaced by its constant part
% $-2/T$.
\par
{\it (iii) Cyclic reduction\/}: Structures which differ
only by cyclic permutation under the trace have to be
identified. This reduces the number of terms drastically.
\par
{\it (iv) Bianchi identities\/}: In a last step one has to
exploit Bianchi identities to reduce the result to a
standard minimal basis. This is of course relevant only for
the gauged case and requires -- in contrast to the items
{\it (i)--(iii)} -- a more detailed analysis.
\par
Let us in the following concentrate on the special case of
a scalar background (cf.\ [12]). Eq.\ (18) simplifies to
$$ \eqalign{
\Gamma[V]
&= \int_0^\infty {{\rm d}T\over T} {\rm e}^{-m^2T}
{\rm tr} \int {\rm d}^dx_0 {\sum_{n=0}^\infty}
{(-1)^n\over n} T \int_0^{\tau_1=T} \!\!\!\!\!\!{\rm d}\tau_2
\int_0^{\tau_2} {\rm d}\tau_3 ... \int_0^{\tau_{n-1}}
{\rm d}\tau_n \cr
&\times \int {\cal D}y \; {\rm e}^{y(\tau_1)\del_{(1)}}
V^{(1)}(x_0) \;...\; {\rm e}^{y(\tau_n)\del_{(n)}}
V^{(n)}(x_0) \exp \Bigl[ - \int_0^T \!\!\!{\rm d}\tau \;
{\dot{y}^2\over 4} \Bigr]\,.
}
\eqno(22)$$
For the Wick contractions only the contraction rule
for exponentials is needed, giving
$$ \eqalign{
\Gamma[V]
=\int_0^\infty {{\rm d}T\over T}&[4\pi T]^{-d/2}
{\rm e}^{-m^2T} {\rm tr} \int {\rm d}^dx_0
{\sum_{n=0}^\infty} {(-1)^n\over n} T \int_0^{\tau_1=T}
\!\!\!\!\!\! {\rm d}\tau_2 \int_0^{\tau_2} {\rm d}\tau_3
\;...\; \int_0^{\tau_{n-1}} \!\!\!\!\!\! {\rm d}\tau_n \cr
&\times {\rm exp} \Bigl[ - T {\sum_{i<k}} G(u_i,u_k)
\del_{(i)} \del_{(k)} \Bigr] V^{(1)}(x_0) ... V^{(n)}(x_0)
\;.\cr }
\eqno(23)$$
Finally, by rescaling to the unit circle, $\tau_i = T u_i
\;(i=1,...,n)$, and using the scaling property of the Green
function, $G(\tau_1,\tau_2) = T G(u_1,u_2)$, one gets
$$ \eqalign{
\Gamma[V]
= \int_0^\infty {{\rm d}T \over T} &[4\pi T]^{-d/2}
{\rm e}^{-m^2T} {\rm tr} \int {\rm d}^dx_0 {\sum_{n=0}^\infty}
{(-T)^n\over n} \int_0^{u_1=1} \!\!\!\!\!\!{\rm d}u_2\int_0^{u_2}
\!\!\!{\rm d}u_3 \,\;... \int_0^{u_{n-1}} \!\!\!\!\!{\rm d}u_n\cr
&\times {\rm exp} \Bigl[ - T {\sum_{i<k}} G(u_i,u_k)\del_{(i)}
\del_{(k)}\Bigr] V^{(1)}(x_0) ... V^{(n)}(x_0) \;.\cr  }
\eqno(24)$$
The inverse mass expansion to some fixed order in T can now be
simply obtained by expanding the exponential, performing a number
of multi-polynomial integrations and identifying terms, which are
equivalent due to cyclic permutations under the trace. The result
has the form
$$
\Gamma[V] =
\int_0^\infty {{\rm d}T\over T} [4\pi T]^{-d/2} {\rm e}^{-m^2T}
{\sum_{n=1}^\infty} (-T)^n {\rm tr}\; O_n\;,
\eqno(25)$$
With a complete computerization of the method a
calculation of the
coefficients up to $O_{11}$ was achieved. We used the algebraic
languages FORM [13] and -- for identifying
cyclic redundancies --
PERL [14]. We did not perform the final $T$-integration, for it
simply yields a $\Gamma$-function for any fixed order in $T$.
Poles of these $\Gamma$-functions correspond to terms of the
effective action which have to be renormalized at one loop for
the dimension of spacetime considered.
\par
In the following the results to ${\cal O}(T^8)$ are quoted,
using the shorthand-notation
$V_{\kappa\lambda}\equiv\del_\kappa\del_\lambda V(x_0)$:
\vfill\eject
$$\eqalign{
 O_1 &= \int {\rm d}x_0 V
\cr\noalign{\vskip8pt}
 O_2 &= {1\over 2!} \int {\rm d}x_0 V^2
\cr\noalign{\vskip8pt}
 O_3 &= {1\over 3!} \int {\rm d}x_0
 \biggl( V^3 + {1\over 2} \Vka \Vka \biggr)
\cr\noalign{\vskip8pt}
 O_4 &= {1\over 4!} \int {\rm d}x_0
 \biggl( V^4 + 2 V \Vka \Vka
 + {1\over5} \Vkala \Vkala \biggr)
\cr\noalign{\vskip8pt}
 O_5 &= {1\over 5!} \int {\rm d}x_0
 \biggl( V^5 + 3 V^2 \Vka\Vka + 2 V\Vka V\Vka + V\Vkala\Vkala
\cr & \qquad\qquad\qquad
 + {5\over 3}\Vka\Vla\Vkala + {1\over 14}\Vkalamu\Vkalamu\biggr)
\cr\noalign{\vskip8pt}
 O_6 &= {1\over6!} \int {\rm d}x_0
 \biggl( V^6 + 4 V^3 \Vka \Vka + 6 V^2 \Vka V \Vka
 + {12\over7} V^2 \Vkala \Vkala
\cr & \qquad
 + {9\over7} V \Vkala V \Vkala + {26\over7} V \Vkala\Vka\Vla
 + {26\over7} V \Vka \Vla \Vkala + {17\over14} \Vka\Vla\Vka\Vla
\cr & \qquad
 + {18\over7} V \Vka \Vkala \Vla + {9\over7} \Vka\Vka\Vla\Vla
 + {3\over7} V \Vkalamu \Vkalamu + \Vmu \Vkala \Vkalamu
\cr & \qquad
 + \Vmu \Vkalamu \Vkala + {11\over21} \Vkala \Vlamu \Vkamu
 + {1\over42} \Vkalamunu \Vkalamunu \biggr)
\cr\noalign{\vskip8pt}
 O_7 &= {1\over7!} \int {\rm d}x_0
 \biggl( V^7 + 5 V^4 \Vka \Vka + 8 V^3 \Vka V \Vka
 + {9\over2} V^2 \Vka V^2 \Vka + {5\over2} V^3 \Vkala \Vkala
\cr & \qquad + {9\over2} V^2 \Vkala V \Vkala
 + 6 V^2 \Vkala \Vka \Vla + 6 V^2 \Vka \Vla \Vkala
 + {7\over2} V^2 \Vka \Vkala \Vla
\cr & \qquad + {17\over2} V \Vka \Vla \Vka \Vla
 + {7\over2} V \Vka \Vka \Vla \Vla
 + {11\over2} V \Vka \Vla \Vla \Vka
 + {11\over2} V \Vka V \Vkala \Vla
\cr & \qquad + {11\over2} V \Vka V \Vla \Vkala
 + {17\over2} V \Vkala V \Vka \Vla
 + {2\over3} V \Vkalamu V \Vkalamu
 + {5\over6} V^2 \Vkalamu \Vkalamu
\cr & \qquad + {5\over2} V \Vkalamu \Vkala \Vmu
 + {17\over6} V \Vkalamu \Vka \Vlamu
 + {5\over3} V \Vkala \Vkalamu \Vmu
 + {17\over6} V \Vkala \Vmu \Vkalamu
\cr & \qquad + {5\over2} V \Vka \Vlamu \Vkalamu
 + {5\over3} V \Vka \Vkalamu \Vlamu
 + {5\over3} \Vka \Vka \Vlamu \Vlamu
 + {11\over6} \Vka \Vlamu \Vka \Vlamu
\cr }$$
\vfill\eject
$$\eqalign{
    & \quad + {35\over9} \Vkalamu \Vka \Vla \Vmu
     + {11\over3} V \Vkala \Vkamu \Vlamu
     + {35\over18} \Vkala \Vka \Vlamu \Vmu
     + {35\over18} \Vkala \Vmu \Vlamu \Vka
\cr & \quad + {97\over18} \Vkala \Vkamu \Vla \Vmu
     + {43\over18} \Vkala \Vkamu \Vmu \Vla
     + {1\over6} V \Vkalamunu \Vkalamunu
     + {1\over2} \Vka \Vlamunu \Vkalamunu
\cr & \quad + {1\over2} \Vka \Vkalamunu \Vlamunu
     + {7\over10} \Vkala \Vmunu \Vkalamunu
     + {16\over15} \Vkala \Vkamunu \Vlamunu
     + {1\over132} \Vkalamunuro \Vkalamunuro \biggr)
\cr\noalign{\vskip8pt}
 O_8 &= {1\over 8!} \int {\rm d}x_0 \biggl(
      V^8 + 12 \Vka V^2 \Vka V^3  + 10 \Vka V \Vka V^4
      + 6 \Vka V \Vka \Vla V \Vla
\cr & + 14 \Vka V \Vkala V^2 \Vla
      + 6 \Vka \Vka V^5
      + { 22 \over 3 } \Vka \Vka V \Vla V \Vla
      + { 20 \over 9 } \Vka \Vka V \Vla \Vla V
\cr & + { 26 \over 3 } \Vka \Vka \Vla V^2 \Vla
      + { 22 \over 3 } \Vka \Vka \Vla V \Vla V
      + { 40 \over 9 } \Vka \Vka \Vla \Vla V^2
      + { 14 \over 3 } \Vka \Vka \Vlamu V \Vlamu
\cr & + { 76 \over 9 } \Vka \Vkala V^3 \Vla
      + { 26 \over 3 } \Vka \Vkala V^2 \Vla V
      + { 22 \over 3 } \Vka \Vkala V \Vla V^2
      + { 86 \over 9 } \Vka \Vkala \Vmu V \Vlamu
\cr & + { 394 \over 45 }\Vka \Vkala \Vmu \Vla \Vmu
      + 6 \Vka \Vkala \Vmu \Vlamu V
      + 6 \Vka \Vkala \Vmu \Vmu \Vla
      + { 16 \over 3 } \Vka \Vkalamu V^2 \Vlamu
\cr & + { 86 \over 9 } \Vka \Vkalamu V \Vla \Vmu
      + { 14 \over 3 } \Vka \Vkalamu V \Vlamu V
      + { 172 \over 99 } \Vka \Vkalamu \Vnu \Vlamunu
\cr & + { 56 \over 33 } \Vka \Vkalamunu V \Vlamunu
      + { 70 \over 9 } \Vka \Vla V \Vka \Vla V
      + { 124 \over 9 } \Vka \Vla \Vka V \Vla V
\cr & + { 112 \over 9 } \Vka \Vla \Vka \Vla V^2
      + 4 \Vkala V^2 \Vkala V^2
      + 12 \Vkala V \Vka V \Vla V
      + 14 \Vkala V \Vka \Vla V^2
\cr & + { 20 \over 3 } \Vkala V \Vkala V^3
      + { 22 \over 3 } \Vkala V \Vkalamu V \Vmu
      + { 40 \over 9 } \Vkala \Vka V^3 \Vla
      + { 22 \over 3 } \Vkala \Vka V^2 \Vla V
\cr & + { 26 \over 3 } \Vkala \Vka V \Vla V^2
      + { 76 \over 9 } \Vkala \Vka \Vla V^3
      + 6 \Vkala \Vka \Vla \Vmu \Vmu
      + { 86 \over 9 } \Vkala \Vka \Vlamu V \Vmu
\cr & + 6 \Vkala \Vka \Vlamu \Vmu V
      + { 394 \over 45 } \Vkala \Vka \Vmu \Vla \Vmu
      + { 238 \over 45 } \Vkala \Vka \Vmu \Vlamu V
\cr & + { 146 \over 45 } \Vkala \Vka \Vmu \Vmu \Vla
      + { 10 \over 3 } \Vkala \Vkala V^4
      + { 12 \over 5 } \Vkala \Vkala V \Vmu \Vmu
      + { 58 \over 15 } \Vkala \Vkala \Vmu V \Vmu
\cr & + { 12 \over 5 } \Vkala \Vkala \Vmu \Vmu V
      + { 20 \over 33 } \Vkala \Vkala \Vmunu \Vmunu
      + { 22 \over 5 } \Vkala \Vkalamu V^2 \Vmu
\cr & + { 58 \over 15 } \Vkala \Vkalamu V \Vmu V
      + { 12 \over 5 } \Vkala\Vkalamu \Vmu V^2
      + { 838 \over 495 } \Vkala \Vkalamu \Vmunu \Vnu
\cr & + { 290 \over 99 } \Vkala \Vkalamu \Vnu \Vmunu
      + { 362 \over 165 } \Vkala \Vkalamunu V \Vmunu
      + { 290 \over 99 } \Vkala \Vkalamunu \Vmu \Vnu
\cr & + { 142 \over 15 } \Vkala \Vkamu V \Vla \Vmu
      + { 128 \over 15 } \Vkala \Vkamu V \Vlamu V
      + { 238 \over 45 } \Vkala \Vkamu V \Vmu \Vla
\cr }$$
\vfill\eject
$$\eqalign{
    & + { 466 \over 45 } \Vkala \Vkamu \Vla V \Vmu
      + { 142 \over 15 } \Vkala \Vkamu \Vla \Vmu V
      + { 92 \over 15 } \Vkala \Vkamu \Vlamu V^2
\cr & + { 334 \over 99 } \Vkala \Vkamu \Vlanu \Vmunu
      + { 146 \over 45 } \Vkala \Vkamu \Vmu V \Vla
      + { 238 \over 45 } \Vkala \Vkamu \Vmu \Vla V
\cr & + { 32 \over 55 } \Vkala \Vkamu \Vmunu \Vlanu
      + { 212 \over 45 } \Vkala \Vkamunu \Vla \Vmunu
      + { 626 \over 45 } \Vkala \Vmu \Vka \Vla \Vmu
\cr & + { 622 \over 45 } \Vkala \Vmu \Vka \Vlamu V
      + { 22 \over 3 } \Vkala \Vmu \Vkala V \Vmu
      + { 22 \over 3 } \Vkala \Vmu \Vkala \Vmu V
\cr & + { 131 \over 165 } \Vkala \Vmunu \Vkala \Vmunu
      + { 22 \over 3 } \Vkalamu V \Vkala \Vmu V
      + { 8 \over 3 } \Vkalamu V \Vkalamu V^2
      + 6 \Vkalamu \Vka V \Vla \Vmu
\cr & + { 14 \over 3 } \Vkalamu \Vka V \Vlamu V
      + 6 \Vkalamu \Vka \Vla V \Vmu
      + { 86 \over 9 } \Vkalamu \Vka \Vla \Vmu V
      + { 16 \over 3 } \Vkalamu \Vka \Vlamu V^2
\cr & + { 172 \over 99 } \Vkalamu \Vka \Vlamunu \Vnu
      + { 412 \over 99 } \Vkalamu \Vka \Vlanu \Vmunu
      + { 12 \over 5 } \Vkalamu \Vkala V^2 \Vmu
      + { 58 \over 15 } \Vkalamu \Vkala V \Vmu V
\cr & + { 22 \over 5 } \Vkalamu \Vkala \Vmu V^2
      + { 290 \over 99 } \Vkalamu \Vkala \Vmunu \Vnu
      + { 838 \over 495 } \Vkalamu \Vkala \Vnu \Vmunu
      + { 4 \over 3 } \Vkalamu\Vkalamu V^3
\cr & + { 10 \over 11 } \Vkalamu \Vkalamu \Vnu \Vnu
      + { 46 \over 33 } \Vkalamu \Vkalamunu V \Vnu
      + { 10 \over 11 } \Vkalamu \Vkalamunu \Vnu V
      + { 2 \over 5 } \Vkalamu \Vkalamunuro \Vnuro
\cr & + { 442 \over 165 } \Vkalamu \Vkalanu V \Vmunu
      + { 2062 \over 495 } \Vkalamu \Vkalanu \Vmu \Vnu
      + { 442 \over 165 } \Vkalamu \Vkalanu \Vmunu V
\cr & + { 838 \over 495 } \Vkalamu \Vkalanu \Vnu \Vmu
      + { 628 \over 99 } \Vkalamu \Vkanu \Vla \Vmunu
      + { 212 \over 45 } \Vkalamu \Vkanu \Vlamu \Vnu
\cr & + { 524 \over 165 } \Vkalamu \Vkanu \Vlamunu V
      + { 412 \over 99 } \Vkalamu \Vkanu \Vlanu \Vmu
      + { 12 \over 11 } \Vkalamu \Vnu \Vkalamu \Vnu
\cr & + { 10 \over 33 } \Vkalamunu V \Vkalamunu V
      + { 344 \over 99 } \Vkalamunu \Vka \Vlamu \Vnu
      + { 56 \over 33 } \Vkalamunu \Vka \Vlamunu V
\cr & + { 40 \over 33 } \Vkalamunu \Vkala V \Vmunu
      + { 290 \over 99 } \Vkalamunu \Vkala \Vmu \Vnu
      + { 362 \over 165 } \Vkalamunu \Vkala \Vmunu V
\cr & + { 10 \over 11 } \Vkalamunu \Vkalamu V \Vnu
      + { 46 \over 33 } \Vkalamunu \Vkalamu \Vnu V
      + { 4 \over 11 } \Vkalamunu \Vkalamunu V^2
\cr & + { 2 \over 9 } \Vkalamunu \Vkalamunuro \Vro
      + { 296 \over 495 } \Vkalamunu \Vkalamuro \Vnuro
      + { 134 \over 165 } \Vkalamunu \Vkalaro \Vmunuro
\cr & + { 2 \over 5 } \Vkalamunuro \Vkalamu \Vnuro
      + { 2 \over 9 } \Vkalamunuro \Vkalamunu \Vro
      + { 2 \over 33 } \Vkalamunuro \Vkalamunuro V
\cr & + { 1 \over 429 } \Vkalamunurosi \Vkalamunurosi
\biggr) \cr }$$
\vfill\eject
The following points about the calculation above should
be emphasized:
\par
{\it (i)} Though the variable $u_1$ has been eliminated, cyclic
invariance has not been broken. All cyclic permutations of a
given term actually occur during the calculation. The polynomial
integrations yield the same factor for terms,
which are equivalent
by cyclic permutation under the trace. This is a
consequence of
the translational invariance of our Green function. Therefore
identification of equivalent terms can easily be done.
\par
{\it (ii)} The final result has a unique minimal form,
which does
not contain any box operators $\del^2$. Since the Green function
$G$ has the property $G(u_i,u_i)=0$, there are
no self-contractions
of exponentials and box operators can never occur. Therefore no
partial integrations with respect to $x_0$ were needed to obtain
the result in a minimal basis.
%%%%%%%%%%%%%%%%%%
The fact that the set of all terms containing no box operators is
a basis for the space of all Lorentz scalars
which can be constructed with $\del$ and V alone is obvious,
as one can always remove box operators by partial integrations.
Furthermore, it is not difficult to see that it is impossible
to construct total derivatives containing no
box operators. This shows the minimality of our basis in the
sense that it does not contain redundant terms (it does {\sl not}
strictly prove that the number of terms making up this basis is,
for any fixed mass dimension, the minimal possible one, though we
consider this to be likely).
\par
Finally, let us try to make clear what is at the root of the gain
in efficiency over standard heat kernel methods. For this purpose,
observe that our choice of the background charge in eq.\ (7) was
a mere matter of convenience; one could have used any periodic
function $\rho$ here fulfilling the constraint
$$
\int_0^T d\tau \rho(\tau ) = 1 .
\eqno(26)$$
To solve eq.\ (7) for general $\rho$, one can first solve it for
the case
$$
\rho (\tau) = \delta (\tau - \sigma)\;,
\eqno(27)$$
leading to
$$
G(\tau_1,\tau_2;\sigma) = | \tau_1 - \tau_2 |
- | \tau_1 - \sigma | - | \tau_2 - \sigma |
+ {2\over T} (\tau_1 -\sigma )(\tau_2 -\sigma )\;,
\eqno(28)$$
and then convolute the result with $\rho$ :
$$
G^{(\rho)}(\tau_1,\tau_2) = \int_0^T d\sigma
\rho (\sigma )G(\tau_1,\tau_2;\sigma)\; .
\eqno(29)$$
Any such $G^{\rho}$ could be used as a Green function for the
evaluation of the path integral, and would lead to a different
effective Lagrangian.
%\eject
\par
To see which choice of $\rho$ corresponds to the standard
heat kernel, we compare with the Feynman--Kac path integral
representation of the heat kernel [15]. This
path integral arises from our eq.\ (5) by a split of the
${\cal D}x$-integral different from eq.\ (13), namely
$$\eqalign{
\int {\cal D}x &= \int {\rm d}^dx_0 \int {\cal D} y \;\cr
x^{\mu}(\tau ) &= x^{\mu}_0 + y^{\mu} (\tau ) \;\cr
y^{\mu} (0) &= y^{\mu} (T) = 0.\cr}
\eqno(30)$$
In this case, the effective Lagrangian ${\cal L}(x_0)$
is therefore obtained as a path integral over the space of all
loops intersecting in $x_0$, instead of having $x_0$ as their
common center of mass. The Green function appropriate to this
boundary condition for
$y^{\mu}$ is
$$
G(\tau_1,\tau_2) = | \tau_1 - \tau_2 |
- \tau_1 - \tau_2 + {2\over T}\tau_1 \tau_2
\eqno(31)$$
corresponding to the choice
$$
\rho(\tau ) = \delta (\tau)\;.
\eqno(32)$$
Following earlier work by Onofri [16], this Green function
was applied by Zuk to the calculation of the higher derivative
expansion both for the ungauged and the gauged case [17] (see
also [18]). However, the resulting
form  of the effective Lagrangian turns out to be
highly redundant,
as has already been discussed in [12].
This is intuitively reasonable, as the split of the path integral
which we have been using is clearly the most ``symmetric'' one.
\par
To summarize, we have applied the ``string-inspired'' method of
evaluating one--loop worldline path integrals to the calculation
of the inverse mass expansion of the one-loop effective action.
The difference between our approach and earlier heat kernel
calculations has been traced to the different boundary conditions
imposed on the path integral. For the
ungauged case, we have completely computerized the method,
and obtained the order ${\cal O}(T^{11})$ of this expansion (the
complete result of this calculation is available on request).
Computerization of the gauged case is in progress. Inclusion of
background gravitational fields is under consideration,
as well as a two--loop generalization based on the construction
of generalized worldline Green functions [19].
\vfill
\eject
\def\litlist#1#2{\hbox to 16cm{\hbox to 10mm {\hfill
    [\hbox to 3.2mm{\hfil #1}]\hskip 3mm}#2\hfill}}
\def\litnonr#1{\hbox to 16cm {\hbox to 10mm {\hfill}#1 \hfill}}
\n{\reffnt References}\par\vskip 3mm
\litlist{1}{R.\ D.\  Ball, Phys.\ Rept.\ {\bf 182} (1989) 1 }
\litnonr{I.\ G.\ Avramidi, Nucl.\ Phys.\ {\bf B 355} (1991) 712}
\litlist{2}{J.\ Schwinger, Phys.\ Rev.\ {\bf 82} (1951) 664}
\litlist{3}{A.\ O.\ Barvinsky, G.\ A.\ Vilkovisky, Nucl.\ Phys.\
    {\bf B 282} (1987) 163;}
\litnonr{Nucl.\ Phys.\ {\bf B 333} (1990) 471;
    Nucl.\ Phys.\ {\bf B 333} (1990) 512}
\litlist{4}{D.\ Ebert, A.\ A.\  Belkov, A.\ V.\ Lanyov,
    A.\ Schaale,}
\litnonr{Int.\ Journ.\ Mod.\ Phys.\ {\bf A 8} (1993) 1313}
\litlist{5}{L.\ Carson, L.\ McLerran, Phys.\ Rev.\ {\bf D 41}
    (1990) 647}
\litnonr{M.\ Hellmund, J.\ Kripfganz, \ M.\ G.\ Schmidt,
HD-THEP-93-23,}
\litnonr{hep-ph/9307284 (to appear in  Phys.\ Rev.\ {\bf D})}
\litlist{6}{L.\ Carson, Phys.\ Rev.\ {\bf D 42} (1990) 2853}
\litlist{7}{A.\ van de Ven, Nucl.\ Phys.\ {\bf B 250}
(1985) 593}
\litlist{8}{Z.\ Bern, D.\ A.\ Kosower, Phys.\
Rev.\ Lett.\ {\bf B 66}
    (1991) 1669;}
\litnonr{Nucl.\ Phys.\ {\bf B 379} (1992) 451}
\litnonr{Z.\ Bern, D.\ C.\ Dunbar, Nucl.\ Phys.\ {\bf B 379}
    (1992) 562}
\litlist{9}{M.\ J.\ Strassler, Nucl.\
Phys.\ {\bf B 385} (1992) 145}
\litlist{10}{A.\ M.\ Polyakov, ``Gauge Fields and Strings'',
     Harwood (1987)}
\litlist{11}{M.\ G.\ Schmidt, C.\ Schubert, Phys.\ Lett.\
     {\bf B 318} (1993) 438,}
\litnonr{hep-th/9309055}
\litlist{12}{D.\ Fliegner, M.\ G.\ Schmidt, C.\ Schubert,
Z.\ Phys.\
     {\bf C 64} (1994) 111,}
\litnonr{hep-ph/9401221}
\litlist{13}{J.\ A.\ M.\ Vermaseren, ``Symbolic Manipulation with
     FORM}
\litnonr{(Version 2)'', CAN (1991)}
\litlist{14}{L.\ Wall, R.\ L.\ Schwartz, ``Programming PERL'',
     O'Reilly (1990)}
\litlist{15}{Y.\ Fujiwara, T.\ A.\ Osborn, S.\ F.\ J.\ Wilk,
     Phys.\ Rev.\ {\bf A 25} (1982) 14}
\litlist{16}{E.\ Onofri, Am.\ J.\ Phys.\ {\bf 46} (1978) 379}
\litlist{17}{J.\ A.\ Zuk, Journ.\ Phys.\ {\bf A 18} (1985) 1795;}
\litnonr{Phys.\ Rev.\ {\bf D 34} (1986) 1791}
\litlist{18}{D.\ G.\ C.\ McKeon, Can.\ J.\ Phys.\ {\bf 70} (1992)
     652}
\litlist{19}{M.\ G.\ Schmidt, C.\ Schubert, Phys.\ Lett.\
     {\bf B 331} (1994) 69,}
\litnonr{hep-th/9403158}
\bye